\documentclass[12pt,preprint]{aastex}

\slugcomment{To appear in AJ}

\shorttitle{Deep H~{\small I} Survey of NGC 2403}
\shortauthors{Fraternali et al.}
\usepackage{graphicx}  

\begin{document}

\title{Deep H~{\small I} Survey of the Spiral Galaxy NGC 2403}

\author{Filippo Fraternali}
\affil{Istituto di Radioastronomia (CNR), via Gobetti 101, 40129, Bologna, Italy}
\affil{Osservatorio Astronomico, Bologna, Italy}
\email{ffratern@ira.bo.cnr.it}

\author{Gustaaf van Moorsel}
\affil{NRAO, P.O.\ Box 0, Socorro, NM 87801, USA}
\email{gvanmoor@aoc.nrao.edu}

\author{Renzo Sancisi}
\affil{Osservatorio Astronomico, via Ranzani 1, 40127, Bologna, Italy}
\affil{Kapteyn Astronomical Institute, University of Groningen, The Netherlands}
\email{sancisi@bo.astro.it}

\author{Tom Oosterloo}
\affil{ASTRON,  P.O.\ Box 2, 7990 AA, Dwingeloo, The Netherlands}
\email{oosterloo@nfra.nl}

\begin{abstract}
High sensitivity H~{\small I} observations of the nearby spiral galaxy
NGC~2403 obtained with the VLA are presented and
discussed. The properties of the extended, differentially rotating
H~{\small I} layer with its H~{\small I} holes, spiral structure and outer
warp are described.  In addition, these new data reveal the presence of a
faint, extended and kinematically anomalous component. 
This shows up in the H~{\small I} line profiles as extended wings of emission
towards the systemic velocity.
In the central regions these wings are very broad (up to 150 km~s$^{-1}$) and
indicate large deviations from circular motion. 
We have separated the anomalous gas component from
the cold disk and have obtained for it a separate velocity
field and a separate rotation curve.  
The mass of the anomalous component is 1/10 of the total H~{\small I} mass.
The rotation velocity of the anomalous gas
is 25$-$50 km~s$^{-1}$ lower than that of the disk.  Its velocity field has
non-orthogonal major and minor axes that we interpret as due to an overall inflow
motion of 10$-$20 km~s$^{-1}$ towards the centre of the galaxy.
   
The picture emerging from these observations is that of a cold H~{\small I}
disk surrounded by a thick and clumpy H~{\small I} layer characterized by
slower rotation and inflow motion towards the center.  The origin of this
anomalous gas layer is unclear. It is
likely, however, that it is related to the high rate of star formation
in the disk of NGC~2403 and that its kinematics is the result of a
galactic fountain type of mechanism. 
We suggest that these anomalous H~{\small I} complexes may be
analogous to a part of the High Velocity Clouds of our Galaxy.

\end{abstract}
 

\section{Introduction}
 
In recent years the study of the vertical structure and kinematics of
the H~{\small I} disks of spiral galaxies has been pursued almost 
exclusively with 
observations of galaxies viewed either edge-on or face-on.
The study of edge-on systems has revealed vertical extensions of
the H~{\small I} layer up to several kpc from the plane and, in the case of
NGC~891, also a slower rotation velocity of the gas in the halo
\citep{swa97}.
In the face-on or nearly face-on galaxies the observations have shown
vertical motions of neutral gas frequently associated with
`holes' in the H~{\small I} distribution \citep{puc92, kam93}.
These results point at a substantial gas circulation between
disk and halo (review by \citet{san99}).
The ejection of gas out of the disk is thought to be related to star formation
in the disk and to be due to stellar winds and supernova explosions that blow
out ionized gas. 
After cooling, this gas falls down to the plane. Such
processes are generally described as {\it galactic fountains}
\citep{sha76, bre80}.  
There are cases, however, in which the vertical motions originate from
external events such as collisions with intergalactic clouds and with small
companions \citep{hul88}.
Accretion of intergalactic `primordial' gas, as 
proposed for the High Velocity Clouds in our Galaxy \citep{oor70}, may also
play a role.

\begin{figure}
\includegraphics[width=\textwidth]{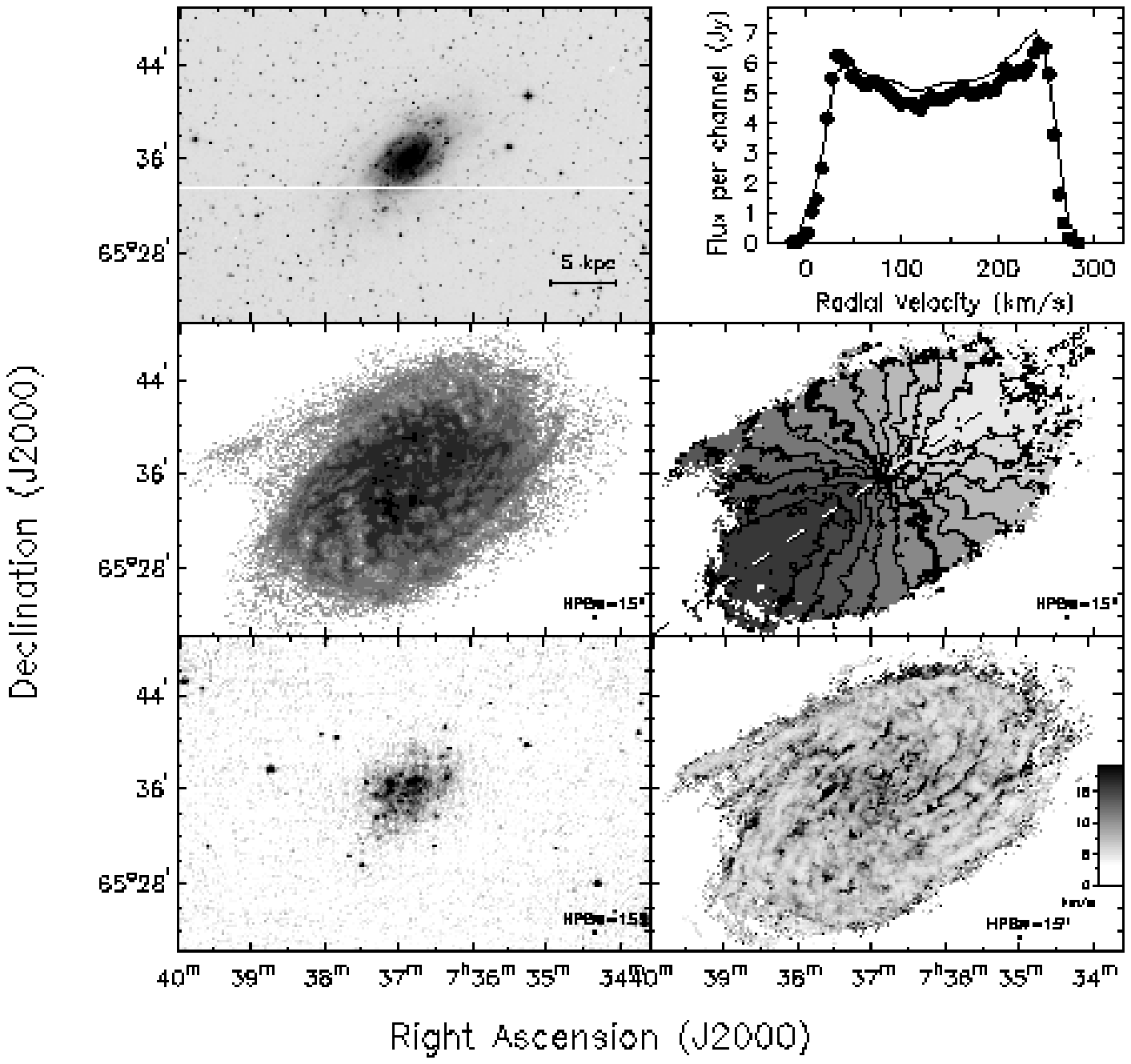}
\figcaption[]{From upper left: optical image (DSS) of NGC~2403,
global H~{\small I} profile (dots) compared with previous single dish 
  observations (line) by Rots (1980), total H~{\small I} map, 
  velocity field, radio continuum, and velocity
dispersion field. These are all at full resolution (15$''$).
The column density in the total H~{\small I} map ranges from
1$\times$10$^{20}$ 
to 5$\times$10$^{21}$ cm$^{-2}$.
In the velocity field (central right panel) the contours are separated by
20 km~s$^{-1}$ and the thicker line marks the systemic velocity (133
km~s$^{-1}$).
The receding side is darker.
\label{fig1}}
\end{figure}

The present study represents a first attempt to use objects at intermediate
inclination angles which offer the advantage that, for a given object,
information is obtained on both the density structure and the kinematics of
H~{\small I} in the vertical direction.  
The nearby spiral galaxy NGC~2403 is a suitable candidate because of
its inclination angle of $\sim$~60$^{\circ}$, and also because of its extended
H~{\small I} layer (about twice the optical size), its regular kinematics and
its symmetric, flat rotation curve.  NGC~2403 is an isolated ScIII
galaxy (Figure~\ref{fig1}) with a bright optical disk of about 8
kpc diameter surrounded by outer, faint-surface-brightness `fossil arms'
\citep{san81}.  It hosts rich H~{\small II} regions with brightnesses
comparable to that of the 30~Doradus complex \citep{dri99}.

Previous H~{\small I} observations with the Westerbork Synthesis Radio
Telescope (WSRT) \citep{beg87,sic97,sch00} of NGC~2403 showed systematic
asymmetries in the H~{\small I} velocity profiles along the major axis in the
form of wings on the side of the lower rotation velocities, towards systemic.
This pattern (we refer to it as `the beard') is similar to that found in
galaxies viewed edge-on or observed with low angular resolution but it is
totally unexpected in NGC~2403 if its H~{\small I} disk is thin.  
Schaap et al. (2000) re-analyzed Sicking's (1997)
observations of NGC~2403 and investigated the effect of the layer thickness in
order to explain such a pattern.  They concluded that the beard may be
produced by H~{\small I} gas located above the plane (in a thick disk or halo)
and rotating more slowly than the disk. Their picture is very similar to that
proposed by Swaters et al. (1997) for NGC 891.

We obtained new H~{\small I} observations of NGC~2403 with the Very
Large Array.\footnote{The National Radio Astronomy Observatory (NRAO) is a
facility of
the National Science Foundation operated under cooperative agreement by
Associated Universities}
These have led to a considerable improvement of the observational
picture and to the detection of new, very faint emission from H~{\small I}
with remarkably anomalous kinematics.

\section{Observations}

We have observed NGC~2403 with the CS configuration of the VLA on four
separate days in January and February, 1999.  This configuration differs from the
regular C-array in that the middle (fifth) telescope on the North arm is
located at the innermost North arm location of the D array.  This
 leads to a spatial resolution similar to that of the regular
C-array, while adding short baselines -- and hence sensitivity at larger
angular scales -- not-found in regular C-array observations \citep{rup98}.
Each observation was 12 hours in duration, of which
approximately 10 hours were spent on source.  The remaining time was used to
observe one phase calibrator (0841+708) and three flux calibrators (3C48,
3C147, and 3C28).  The latter were used for bandpass calibration as well.
The observational parameters are summarized in Table~\ref{tab1}.

\subsection{Data reduction}

Calibration and editing of the data were done separately for each of the four
data-sets, using standard procedures of the AIPS (Astronomical Image
Processing System) package. Nineteen channels at both ends of the band
(excluding the very edges) were identified as being free of line emission.
The AIPS task UVLIN \citep{cor92} was then used to interpolate this continuum
emission across the range of channels containing line emission, and subtract
it from the $uv$ data.

The resulting continuum-free $uv$ dataset containing the combined data for all
four days was then further reduced using the MIRIAD (Multichannel Image
Reconstruction, Image Analysis and Display) package.  The data cube was made
using Robust or Briggs' weighting \citep{bri95} with robustness 0.2 getting a
spatial resolution of $\sim$~15$''$ or 230 pc at the adopted distance 
of 3.18 Mpc \citep{mad91}.
Another data cube was made by applying a Gaussian taper in the visibility
domain.  This data cube has a spatial resolution of $\sim$~30$''$.  The two
cubes have 62 channels ranging in velocity from $-$24.7 km~s$^{-1}$ to 289.6
km~s$^{-1}$ and spaced by 5.15 km~s$^{-1}$. 
To these data a Hanning smoothing
was applied in velocity, leading to a velocity resolution of 10.3 km~s$^{-1}$.
The dirty images were deconvolved using a Clark CLEAN algorithm \citep{cla80}.
To identify the region of emission, the cubes were firstly CLEANed with a
cutoff of 3 times the r.m.s.\ noise.  The resulting images were then used to
interactively define the regions of line emission in each channel.  Next, a
deep CLEAN (cutoff of $\sim$~0.2 $\sigma$) was performed.

\begin{figure}
\plotone{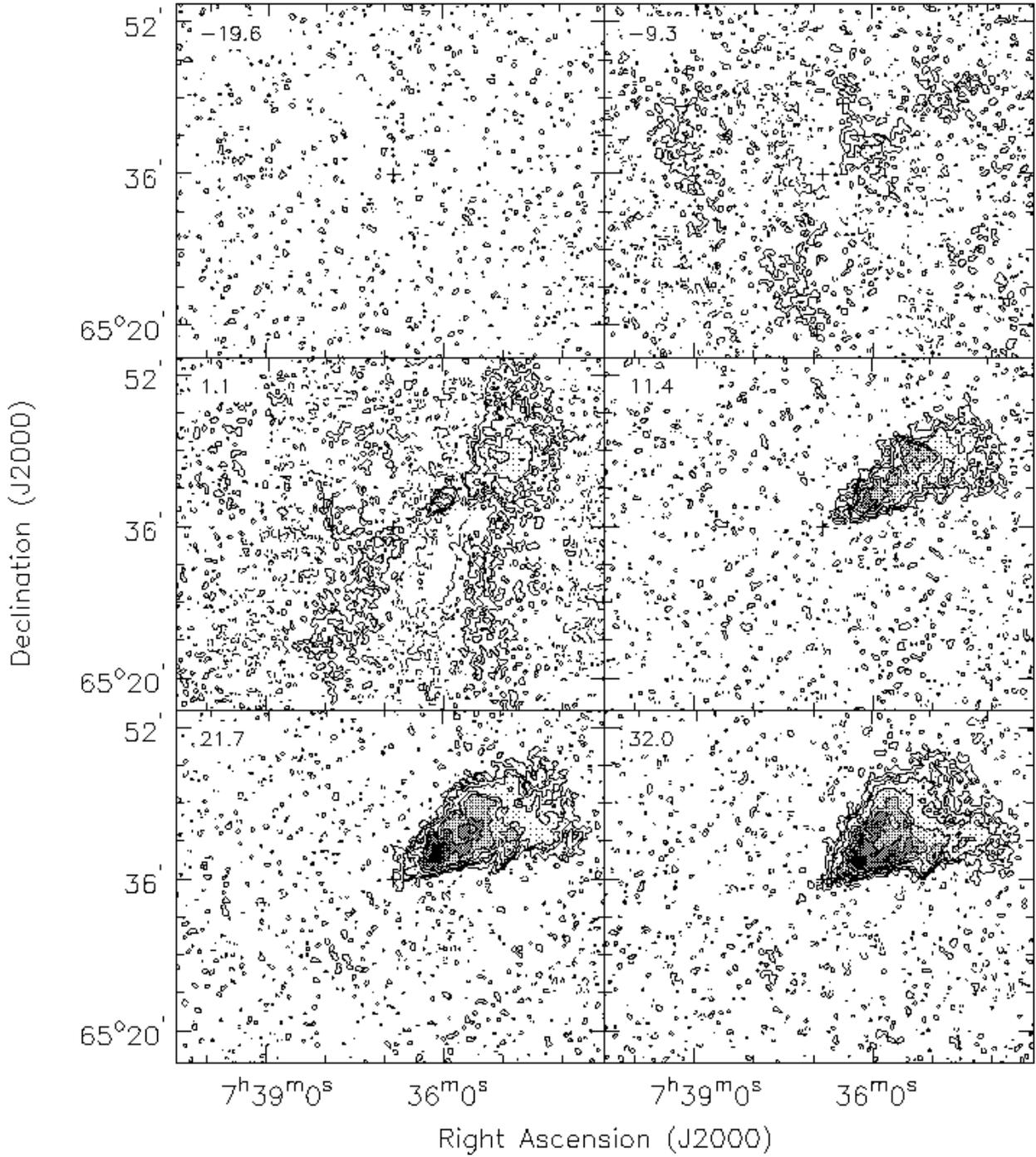}
\figcaption[]{H~{\small I} channel maps at 30$''$ resolution (Hanning smoothed).
The heliocentric radial velocities (km~s$^{-1}$) are shown in the upper 
left corners. The
contours are $-$0.5, 0.5, 1, 2, 5, 10, 20, 50 mJy/beam;
the r.m.s.\ noise is 0.22 mJy/beam.
The cross indicates the kinematical
centre of the galaxy.
\label{fig2}}
\end{figure}

After cleaning, a few
channel maps still showed some spurious effects (Figure~\ref{fig2}, 
top left corner of maps from 217.5 km~s$^{-1}$ to 248.4
km~s$^{-1}$). These may indicate the presence of very extended H~{\small I}
associated with the outer parts of the warp. 
Some channel maps show serious
contamination from H~{\small I} emission from our Galaxy at velocities around
0 km~s$^{-1}$ (Figure~\ref{fig2}, see the channel maps
between $\sim$~$-$10 km~s$^{-1}$ and $\sim$~$+$10 km~s$^{-1}$).  
This contamination is slightly reduced by applying the Hanning smoothing and
also by using high resolution data because it generally has a greater angular
scale than the emission from NGC~2403.

Comparison of our global H~{\small I} profile with a previous single dish
profile from \citet{rot80} (Figure~\ref{fig1}) shows we are not missing
significant H~{\small I} flux due to missing short spacings.  
The final data cube at $\sim$~15$''$ resolution has a r.m.s.\ noise per channel
of 0.17 mJy beam$^{-1}$, while the lower resolution data cube has a
r.m.s.\ noise of 0.22 mJy beam$^{-1}$. 
The minimum detectable column density (5 $\sigma$ detection
in one velocity resolution element in the 30$''$ resolution cube)
is $ 2.0 \times 10 ^{19}$ cm$^{-2}$ (0.15 $M_{\odot}$ pc$^{-2}$) corresponding
to $\sim$~2.6 $\times$ 10$^4$ $M_{\odot}$ beam$^{-1}$. This very high
sensitivity has permitted us to detect very faint components in the neutral
hydrogen that were not seen in previous observations.  A summary of the
parameters for the two cubes is given in Table~\ref{tab2}.

\subsection{Data description}

Once the data cubes were obtained, the subsequent analysis was done using the
GIPSY (Groningen Image Processing SYstem) package.  Figure~\ref{fig2} shows
every second channel map of the 30$''$ resolution cube of NGC~2403. We
present here the smoothed data because they show more clearly the faint
H~{\small I} components. Some representative channels at high (15$''$)
resolution are shown in Figure~\ref{fig8} overlaid on an optical DSS image.
We also present several position-velocity (p-v) diagrams parallel to the major
(Figure~\ref{fig4}) and minor (Figure~\ref{fig5}) axes. The cuts are shown in
Figure~\ref{fig3} overlaid on the total H~{\small I} map.

The full resolution total H~{\small I} map (Figure~\ref{fig1}) 
was obtained by adding the channel maps containing H~{\small I} 
emission (from $-$19.6 km~s$^{-1}$ to 279.3 km~s$^{-1}$).
For each channel map the areas with H~{\small I} emission were outlined 
using masks made on the 30$''$ resolution smoothed data cube.
The total H~{\small I} mass, corrected for the primary beam attenuation,
is 3.24 $\pm$ 0.05 $\times$ 10$^9$ $M_{\odot}$.
The value of the mass derived for this galaxy by \citet{beg87} was
3.0 $\times$ 10$^9$ $M_{\odot}$ while \citet{sic97} with more sensitive
observations derived 3.13 $\times$ 10$^9$ $M_{\odot}$ assuming the
distance of 3.18 Mpc.
For comparison the single dish mass \citep{rot80} is 3.31 $\times$ 10$^9$
$M_{\odot}$.

To first order, the distribution and the kinematics of the H~{\small I} are
regular.
Figure~\ref{fig6} (left panel) shows the total H~{\small I} at low (60$''$)
resolution overlaid on the DSS optical image.  The radius of the H~{\small I}
disk down to a column density of $\sim$~0.2 $M_{\odot}$pc$^{-2}$ (2.5
$\times$ 10$^{19}$ cm$^{-2}$) is about 22 kpc (the Holmberg radius is 13 kpc).
The H~{\small I} disk appears to be full of holes, some having kpc sizes
\citep{mas99}.  The very outer layer is warped.  The warp is highly
asymmetric, more pronounced on the eastern side of the galaxy (see channels
from 176.2 km~s$^{-1}$ to 227.8 km~s$^{-1}$) than on the western side where it
is just faintly visible (channels from 42.3 km~s$^{-1}$ to 62.9 km~s$^{-1}$).
The warp is also visible in the total H~{\small I} map and in the velocity
field at low resolution shown in Figure~\ref{fig6} where the H~{\small I} disk
on the N-E side appears to be more extended.  A summary of the optical and
H~{\small I} parameters for NGC~2403 is given in Table~\ref{tab3}.

\subsubsection{The anomalous gas}

The most striking characteristics of the kinematics of the neutral hydrogen in
NGC~2403 are illustrated in Figure~\ref{fig7}. 
This diagram offers a display
of the H~{\small I} emission along the major axis of NGC~2403.  
It was obtained by integrating the 15$''$ full resolution data over a slice
1$'$ wide in order to improve the signal to noise ratio.  
Clearly, the H~{\small I} line profiles are strongly asymmetric with respect
to the peak of the line emission. 
Note that the white squares, which mark the rotation curve, follow
closely the ridge of the H~{\small I}.  
There are H~{\small I} tails extending (as a beard) towards the systemic
velocity. 
The brighter of these tails had been seen in previous 21-cm
observations with the WSRT and already reported by Schaap et al. (2000).
Because of the higher sensitivity of the present observations the tails are now
much easier to recognize and they appear to extend over a much larger region
of the position-velocity map. 
A clearer picture of the extent, the structure
and the kinematics of the gas in the beard is now emerging.  
The density distribution is characterized by sub-structures in the form of
filaments and spurs.  
These are observed all over the H~{\small I} disk at velocities that
differ tens of km~s$^{-1}$ from rotation.  
The most remarkable filament has a size of $\sim$~8 kpc and is visible in the
channel maps between 93.8 km~s$^{-1}$ and 155.6 km~s$^{-1}$.  
It has a coherent structure and kinematics and its total H~{\small I} mass is
$\sim$~1 $\times$ 10$^7$ $M_{\odot}$.  
It does not show any obvious relation with any feature in the optical disk
(see upper right and bottom left panels of Figure~\ref{fig8}) or with the
H~{\small I} spiral arms.
The $-1'$ and $-2'$ p-v plots parallel to the major axis (and
partly also that along the major axis itself) in Figure~\ref{fig4} show this filament
elongated on the low velocity side between 0$'$ and $-$8$'$.  Remarkably, it
follows the overall pattern of the low-density gas suggesting that it is just
a prominent condensation of gas within it and not a separate feature.

The p-v plots parallel to the minor axis (Figure~\ref{fig5}) show an
asymmetric distribution of the low-level emission (tracing the beard).  In the
diagrams from $-$4$'$ to $-$1$'$, at velocities below the systemic velocity
(horizontal line at 133 km~s$^{-1}$), this low-level emission appears
preferentially in the left part of the diagrams, between 0$'$ and $\sim$~$+$10$'$
(North-East side of the galaxy), whereas at velocities above systemic it is
more visible in the right part of the plots. 
In the diagrams from $+1'$ to $+4'$, at velocities above systemic the
low-level emission is more
prominent on the right part of the diagrams; at velocities below systemic it is
in the left part. 
A similar asymmetric distribution
of the low-level H~{\small I} emission is also seen in the channel maps close
to the systemic velocity (Figure~\ref{fig2}) where the lower contours do not
follow the H~{\small I} ridge and seem displaced counter-clockwise.  This
effect indicates a rotation 
of the kinematical axes of the faint emission with respect to the
axes of the cold disk.  As we will discuss later (see Section 3.2.1) such a
change in the kinematical position angles can be attributed to an overall
radial motion of the anomalous gas of about 10$-$20 km~s$^{-1}$ towards the
centre of the galaxy.

Apart from the extended low-level emission described above, the p-v diagrams
in Figure~\ref{fig4} and Figure~\ref{fig7} also show emission in the forbidden
quadrants.  This is particularly clear in the upper right quadrant in
Figure~\ref{fig7} (0$'$ to $-4'$ North-West with respect to the centre) along
the major axis at projected velocities (around 200 km~s$^{-1}$) which differ
from rotation by more than 130 km~s$^{-1}$.  The spatial location of this gas
can be seen in the channel maps between 165.9 km~s$^{-1}$ and 217.5
km~s$^{-1}$ where it is visible just North-West of the centre.  The total mass
of this gas is 4.3 $\times$ 10$^6$ $M_{\odot}$.  The channel maps between 62.9
km~s$^{-1}$ and 104.1 km~s$^{-1}$ show a faint counterpart on the South-East
side of the galaxy with a total mass of 1.7 $\times$ 10$^6$ $M_{\odot}$.  All
this gas at extreme anomalous velocities (the `forbidden' gas) seems to be
confined to the region corresponding to the bright optical part of NGC~2403
(see Figure~\ref{fig8}, top left and bottom right panels).  Its overall kinematical
pattern, as it appears in Figure~\ref{fig7}, indicates that all the anomalous
gas, including the forbidden one, forms one coherent structure, which closely
follows the rotation of the system.

In the following we refer to all the gas at projected anomalous 
  velocities with respect to
the rotation curve (beard + forbidden gas) as the `anomalous gas'.

\section{The separation of the anomalous gas}

We have attempted to separate the anomalous from the regular H~{\small I} assuming that
the H~{\small I} profiles are formed by two components: 1) a narrow one of
Gaussian shape centred on the rotation velocity of the galaxy (the cold disk), and
2) a broader one, mainly at lower rotation velocities and with unknown profile
shape (the anomalous gas).
We have modelled the cold disk by fitting a
Gaussian function to the line profile after clipping the data at 30\% 
of the peak.
With this clipping we fit only the upper part of the
profiles. 
The differences between the Gaussian fit with and without the
clipping are about 1 km~s$^{-1}$ for the value of the rotation velocity and
about 3 km~s$^{-1}$ for the velocity dispersion.  
The velocity field and the distribution of the velocity dispersion of the cold
disk (Figure~\ref{fig1}), obtained with this Gaussian fit, have then
been used to generate a `Gaussian cube' that has been subtracted from the data
leaving the residual anomalous gas.  
A second method has been tried to generate the Gaussian cube
for the cold disk by fitting half a Gaussian to the high
velocity side of the line profiles. 
This is based on the assumption that the anomalous gas does not contribute to
the high velocity side of the line profile. 
The obtained half-Gaussian has then been folded on the low velocity side
of the profile and subtracted from the data. 
The results obtained with the two methods are very similar.

\subsection{The cold disk}

The velocity field and the map of the velocity dispersion of the cold disk are
shown in Figure~\ref{fig1}. 
The velocity field is regular and symmetric.
The velocity dispersion has average values
around 9$-$10 km~s$^{-1}$ except in the central region and in the spiral arms
where it reaches values of up to 10$-$15 km~s$^{-1}$ (darker grayscale). 
After correction for the instrumental resolution these values are about 1
km~s$^{-1}$ lower.
In order to determine the kinematical parameters and the rotation curve of 
NGC~2403 we have performed the standard tilted ring fit of the velocity field
\citep{beg87}. 
The rings have been fitted with a radial increment of 15$''$.  Points were
weighted by the cosine of the azimuthal angle with respect to the major axis.
In the determination of the central position and the systemic velocity only
the rings with radii smaller than 13 kpc were considered.  
The values derived for the position of the centre and for the systemic
velocity are reported in Table~\ref{tab3}. 
While keeping these parameters fixed, we proceeded to fit the position and
inclination angles.  
The position angle (Figure~\ref{fig9}) shows a clear
deviation from the mean value in the central region between 2 and 4 kpc and
remains constant out to about 15 kpc where it starts decreasing in the region
of the warp.  
The inclination angle shows variations in the central region
around a mean value of about 61$^{\circ}$.  
Then it rises to more than 65$^{\circ}$ beyond 15 kpc.  
The mean (error weighted) values for the position
and the inclination angles were derived between
4 and 15 kpc and resulted in 124.5 $\pm$ 0.6$^{\circ}$ and 62.9 $\pm$
2.1$^{\circ}$ respectively.  
The values obtained by \citet{sic97} were 123.9 and 60.9
degrees respectively while \citet{beg87} found 122.5 and 60.2.  
The lower values for the inclination angle found by these authors
agree with the value we get considering only the inner regions (radii $\leq$
10 kpc) of the disk.

While keeping the other parameters fixed, the rotation curve was
derived separately for the approaching and receding side of the galaxy, as
shown in the top left panel of Figure~\ref{fig9}. 
The line shows the rotation curve obtained by \citet{beg87}.
Points beyond 20 kpc were determined using the velocity field
at 30$''$ resolution.
Because of the large deviations of position and inclination angles from
their mean values beyond 20 kpc they were left free to vary in the fit.
The rotation curves for the two sides indicate a good overall symmetry
except for the regions beyond 16 kpc where the approaching side has
velocities 5$-$10 km~s$^{-1}$ higher than the receding one.

Figure~\ref{fig9} also shows the radial column density fitted with an
exponential law for radii larger than 4 kpc (the inferred H~{\small I}
scalelength is 5.7 $\pm$ 0.1 kpc), and the velocity dispersion derived with
the formula ${\sigma_\mathrm{disp}}^2={\sigma_\mathrm{obs}}
^2+{\sigma_\mathrm{instr}}^2$ where
${\sigma_\mathrm{instr}}$ is the instrumental velocity resolution
($\simeq$~2 $\times$ channel separation) and ${\sigma_\mathrm{obs}}$ is the
$\sigma$ of the Gaussian function fitted to the clipped line profile.  The
derived velocity dispersion shows a decrease from the central values of
$\sim$~12 km~s$^{-1}$ to values of $\sim$~8 km~s$^{-1}$ in the outer regions.

\subsubsection{The mass model}

The rotation curve of NGC~2403 was derived again by applying the tilted
ring fitting to the velocity field of the whole galaxy.
For points beyond 20 kpc we again used the smoothed data cube with no
fixed position and inclination angles.
As an estimate of the errors we took the difference in velocity
between the approaching and receding side.
We have subsequently performed the standard analysis of the rotation
curve  with the decomposition into three
mass components \citep{beg89}: the
stellar disk, the gas disk, and the dark matter halo.
No stellar bulge is visible in the optical luminosity profile
\citep{ken87}.
For the stellar disk we have taken the brightness profile published by
\citet{ken87} while for the H~{\small I} profile we multiplied our data by a
scale factor of 1.4 to take into account the Helium abundance.
We assumed a value of 0.4 kpc for the scale height of the stellar
layer \citep{kru81} and 0.2 kpc for gas layer
\citep{sic97}.
The dark matter halo was modelled with the spherically
symmetric density distribution
$\rho(R)=\rho_0(1+\frac{R}{R_0})^{-2}$,
where $\rho_0$ is the central density and $R_0$ is the core radius.

In Figure~\ref{fig10} we show four different fits.
The maximum disk model gives a scaling factor 
($M/L$) for the stellar
component of 2.3, the `Bottema disk' model \citep{bot93} has a fixed $M/L$ of
1.4, and the third model is the so-called maximum halo (no stellar disk).
The fourth panel shows a model with no dark matter halo and a free scale
factor for stars and H~{\small I};
we obtained an $M/L$ of 2.3 and a value of 9.9 for the H~{\small I} 
scaling factor, similar to that found by \citet{hoe01}.
The first three fits are comparable.
Note that the last model gives a higher 
discrepancy in the outer regions than found by Hoekstra et al.\ 2001.
However, this latter produces the best fit of the
inner substructures of the rotation curve.
The present rotation curve is more extended (about 2 kpc) than that by
\citet{beg87} and does not show any sign of declining.
The total dynamical mass inferred within a galactic radius of 22.5 kpc
is 9.4 $\pm$ 0.7 $\times$ 10$^{10}$ $M_{\odot}$. 
This value is consistent with that obtained by \citet{beg87}
(8.4 $\pm$ 0.4 $\times$ 10$^{10}$ $M_{\odot}$) for a galactic radius of 20 kpc.

\subsection{The anomalous gas}

Figure~\ref{fig11} shows the p-v diagram for the anomalous gas along the major
axis of NGC~2403 after the subtraction of the cold disk.
The filled squares show the rotation curve derived for the cold disk.
Figure~\ref{fig12} shows the distribution of the anomalous gas overlaid on the
optical picture (left panel).
Its outer radius is about 15$-$16 kpc and the total H~{\small I} mass is
$\sim$~3 $\times$ 10$^8$ $M_{\odot}$
(1/10 of the total H~{\small I} mass of NGC~2403, 0.3\%
of the total dynamical mass).
For the velocity dispersion of the anomalous gas the data indicate
values of 20$-$50 km~s$^{-1}$.

Where is this anomalous gas located in space with respect to the cold disk?
Is it located above the disk, in the halo region? 
The vertical extent of this gas layer is difficult to determine precisely because
the inclination of the galaxy is not sufficiently high.
Schaap et al. (2000) using a 3D modelling of the
gas layer arrived at possible values for its thickness between 1 and 3 kpc (FWHM).
Similar values were found for the edge-on galaxy NGC~891
(Swaters et al. 1997) whose H~{\small I} `halo' is probably a component
similar to the anomalous gas in NGC~2403.
A similar modelling performed by us 
gives an upper limit for the thickness of the anomalous gas of about 3 kpc, while
the thickness of the disk is presumably lower than 0.4 kpc \citep{sic97}.  
Also from a dynamical point of view this separation into two systems of
different thicknesses is natural as the anomalous gas and the cold
disk, being collisional systems with separate kinematics, can not spatially
co-exist in the same thin layer.

\subsubsection{Velocity field and radial motions}

Figure~\ref{fig12} (right panel) shows the intensity-weighted velocity field
of the anomalous gas. Clearly the kinematics is dominated, as for the cold
disk, by differential rotation. 
The projected kinematical axes, however, appear to be rotated
counter-clockwise with respect to those of the cold disk.  
This is more evident for the minor axis (thick line).  
Furthermore, the two axes are not orthogonal. 
This is a result of the asymmetric distribution of the anomalous
gas already noted in the description of the p-v parallel to the minor axis of
the galaxy (Section 2.2.1).  A simple explanation is that of an overall inflow
of the anomalous gas towards the center of the galaxy \citep{fra01}.

We have explored the possibility of such a radial flow quantitatively using a
tilted ring fit to the velocity field of the anomalous gas.  
This was done by keeping the centre of the galaxy and the systemic velocity
fixed to the values found for the cold disk. 
The resulting inclination angle is similar to that of the disk, but the
position angle, as expected from the counter-clockwise rotation of kinematical
axes, turns out to be significantly higher ($\sim$~131$^{\circ}$).  
Assuming such a rotation of the kinematical axes is indeed due to
radial motions, we have included in the tilted ring
model a radial component for the gas velocity.  
In order to decide about the sign of such radial motion $-$inflow or
outflow$-$ it is necessary to know which is the near and which is the far side
of the galaxy. 
We have taken the South-West side as the near side of NGC~2403.  
This is based on the assumption that the spiral arms are trailing and is also
supported by the presence of dust absorption features (see optical images in
Sandage et al.\ 1981).
Keeping the position angle fixed to the value found for the cold disk (124.5),
the fit gives a radial flow velocity of $-$13.6 $\pm$ 1.2 km~s$^{-1}$ (the
minus sign means inflow) from 4 to 15 kpc.
 
Is it possible that also the cold disk has a non-zero radial
velocity?
We tested this possibility by considering both radial velocity and position
angle as free parameters in the tilted ring fit to the velocity field of the
cold disk.
Figure~\ref{fig13} (right panel) shows the trend of radial velocity
for both the disk and the anomalous gas obtained with this method.
The mean values of radial velocity between 4 and 15 kpc are $-$3.1 $\pm$
0.3 km~s$^{-1}$ and $-$16.2 $\pm$ 1.1 km~s$^{-1}$ for the disk and the
anomalous gas respectively. 
Thus the velocity field of the cold disk also shows a small distortion that
could be explained as a weak radial inflow.
However, an alternative and more likely interpretation for this distortion
is that of triaxiality of the potential.
Indeed, the deviation from circular motion found here is comparable 
to that found by \citet{sch97} (about 2 km~s$^{-1}$) with a Fourier
analysis of the orbits.
Such a weak triaxiality would, instead, be totally insufficient to account 
for the effect (a factor of 10 greater) observed for the anomalous gas.

Figure~\ref{fig13} also shows the rotation curve of the disk compared with
the mean rotation velocity of the anomalous gas.
The difference is about 25 km~s$^{-1}$ in the outer parts, increasing 
to 40$-$50 km~s$^{-1}$ in the inner regions.
Table~\ref{tab4} summarizes the properties of the H~{\small I} disk and of the 
anomalous gas for NGC~2403.

  \section{Summary and concluding remarks}

  We have presented new, deep H~{\small I} observations of the spiral 
  galaxy NGC~2403. These have been used to study the structure of the 
  gas layer and the dynamics of the system.
  The main results are briefly summarized here.

  A new faint, extended gas component with anomalous velocities has 
  been detected. This H~{\small I} component (the `anomalous gas') shows up 
  in the line profiles as broad wings at velocities lower than rotation 
  and partly also at forbidden (non-circular) velocities. 
  Such wings are not explained by inclination effects, pure thickness or
  poor angular resolution, but result from the presence of a  
  thick layer of H~{\small I} which rotates more slowly than the gas 
  in the plane (see also Schaap et al. 2000).
 
  We have separated the anomalous gas from the thin disk of cold 
  H~{\small I} assuming a Gaussian profile for the latter.
  The anomalous H~{\small I} has a mass of about 3~$\times$ 10$^8$
  $M_{\odot}$ (1/10 of the total H~{\small I} mass), a mean rotation velocity
  25$-$50 km~s$^{-1}$ lower than that of the disk, and a radial 
  inflow of 10$-$20 km~s$^{-1}$.
  
  The origin of this gas is still a matter of debate. The main issue is 
  whether it is the result of processes taking place in the system itself
  or whether it could be infall of extragalactic, possibly primordial gas 
  (cf.\ Oort 1970).
  The overall kinematical pattern of the anomalous gas, its highly coherent 
  structure and its close connection with the disk of NGC 2403 seem to 
  point at an internal origin. 
  Indeed, there may be a close relation with 
  the process of active star formation going on in the disk of NGC 2403 
  and a galactic fountain type of mechanism may provide the explanation 
  for both its origin and its dynamics.
  However, the detection of H~{\small I} at forbidden velocities (more 
  than 130 km~s$^{-1}$ from rotation) and the presence of coherent structures
  with sizes up to 8 kpc, are puzzling.
  Considering the large velocity dispersion (about 20$-$50 km~s$^{-1}$) and
  the lower 
  rotation velocities found for the anomalous H~{\small I}, it is tempting to 
  describe 
  what we see here in terms of a large asymmetric drift of a thick and clumpy
  gas layer.
  
  The results of the present H~{\small I} observations of NGC~2403 have 
  also interesting implications regarding the High Velocity Clouds 
  observed in the Milky Way \citep{wak97} whose distances,
  nature and origin are still a matter of debate.
  It is possible that the anomalous H~{\small I} complexes discovered
  in NGC~2403 are analogous to the HVCs seen in the Galaxy, or at least
to a class of them.
Consider for instance the 8~kpc filament, the largest structure found in 
NGC~2403, with a projected velocity
difference from the rotation of 60$-$100 km~s$^{-1}$ and an H~{\small I} mass of about 1 
$\times$ 10$^7$ $M_{\odot}$.
If viewed from inside NGC~2403 it would probably appear to 
have a large angular size similar
to that of some extended HVC complexes e.g.\ the well known complex C \citep{wak99}.
To be as massive as the filament in NGC~2403, complex C would have to be
as far as 15 kpc from us, compatible with the observational lower limits 
\citep{wak97}.
Such a distance would also be of the same order as the distance (between
4 and 10 kpc) estimated for complex A \citep{woe99} that shows several similarities 
with complex C.
  The fact that NGC~2403 is a normal, non-interacting spiral galaxy 
  suggests that the H~{\small I} features discovered here may be common 
  among spiral galaxies and that, perhaps, they have not been detected yet 
  because of the low sensitivity of previous surveys.
 
  Apart from the investigation of the newly discovered anomalous 
  gas component, the present observations have also been used for the study 
  of the dynamics of NGC 2403. The main results are the finding of an 
  extended, highly asymmetrical warp of the H~{\small I} outer layer and the 
  derivation of a new rotation curve. This has been analyzed with the 
  standard decomposition technique and the main results on the disk and the 
  dark matter halo are similar to those already obtained by
  \citet{beg87}.

\begin{acknowledgements}
We acknowledge financial support 
from the Italian Ministry for the University and Scientific 
Research (MURST).
\end{acknowledgements}

\clearpage

\clearpage

\clearpage
\begin{figure}
\includegraphics[width=\textwidth]{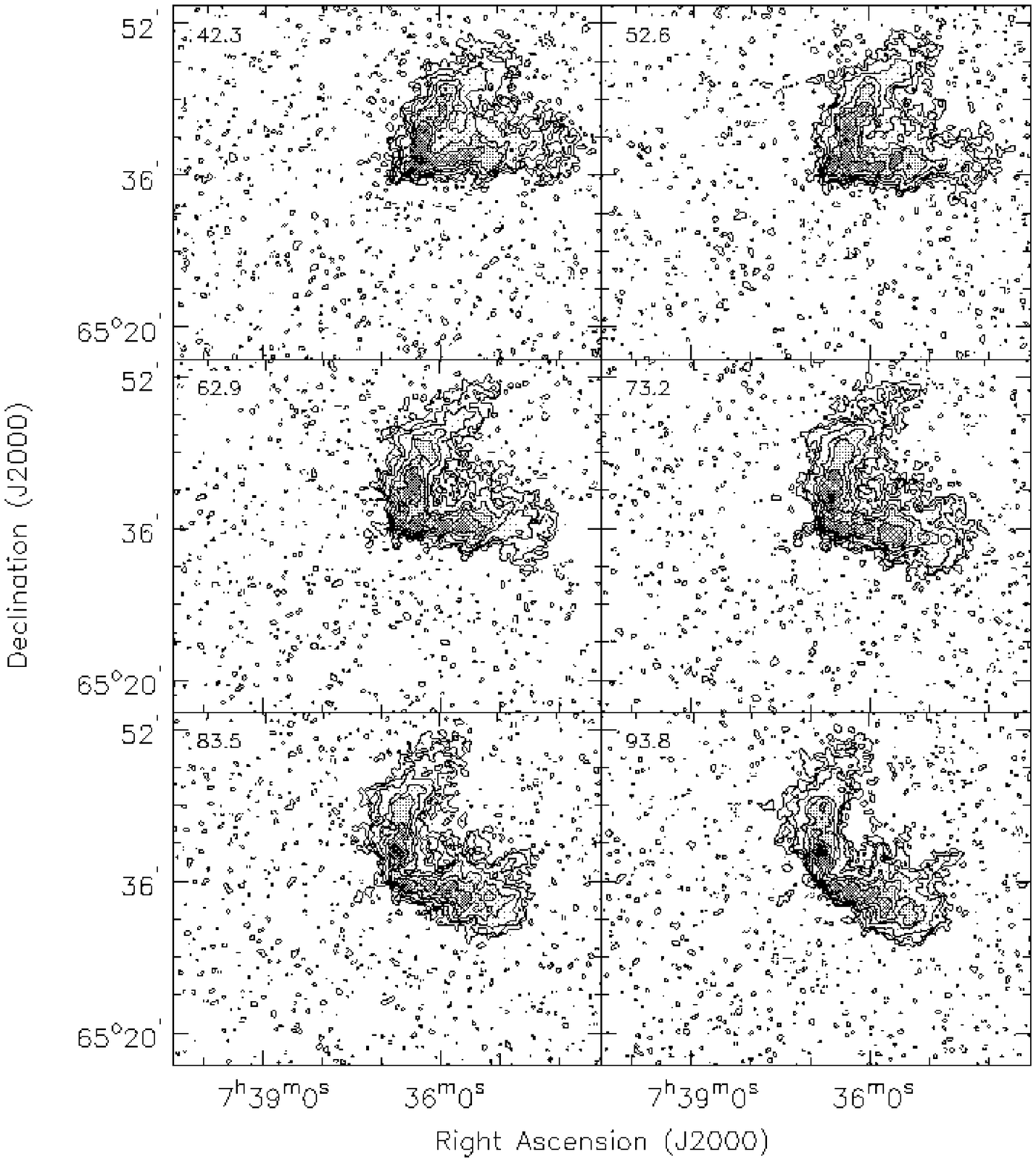}
Figure 2.2 (continue)
\end{figure}

\clearpage
\begin{figure}
\includegraphics[width=\textwidth]{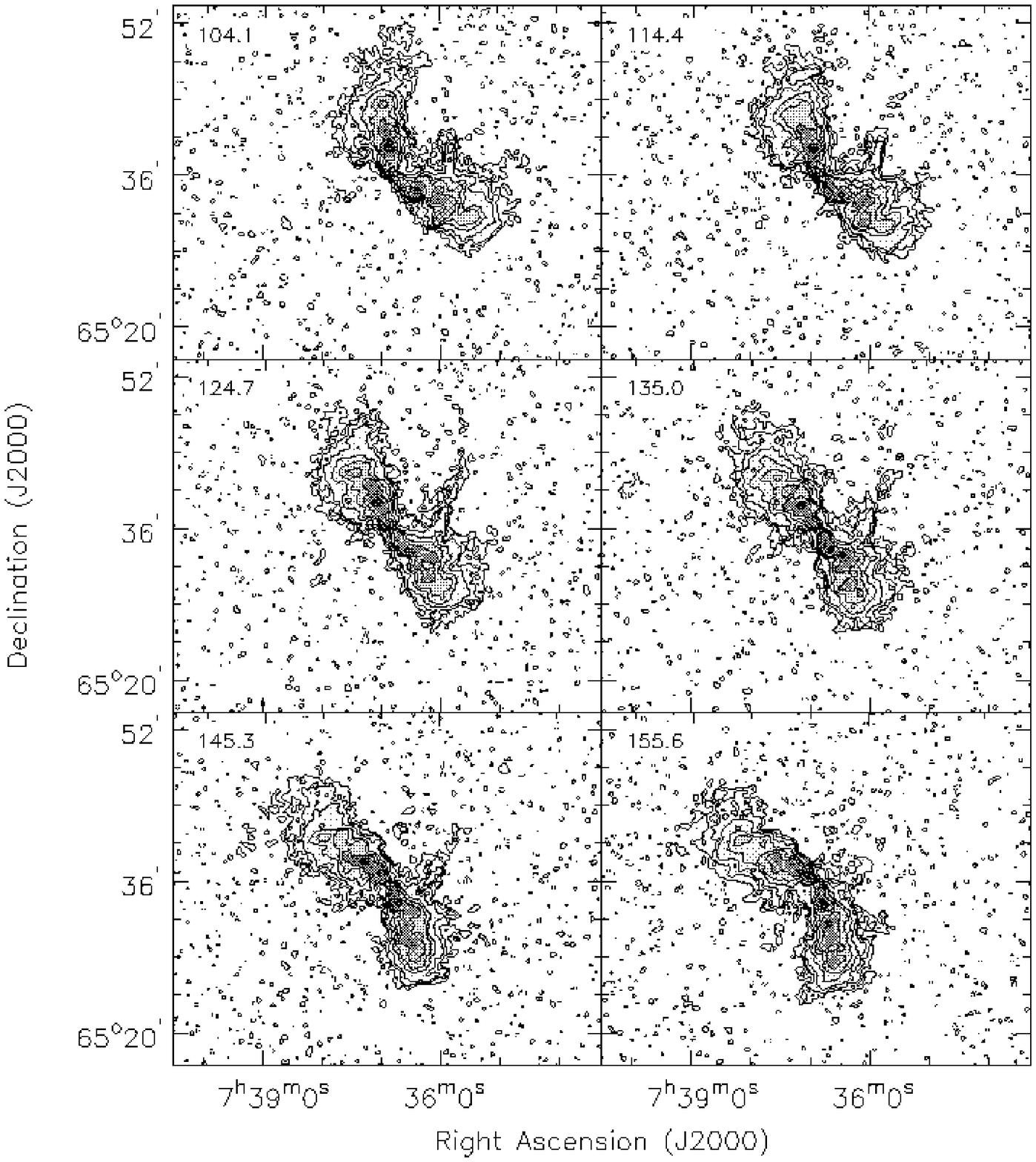}
Figure 2.2 (continue)
\end{figure}

\clearpage
\begin{figure}
\includegraphics[width=\textwidth]{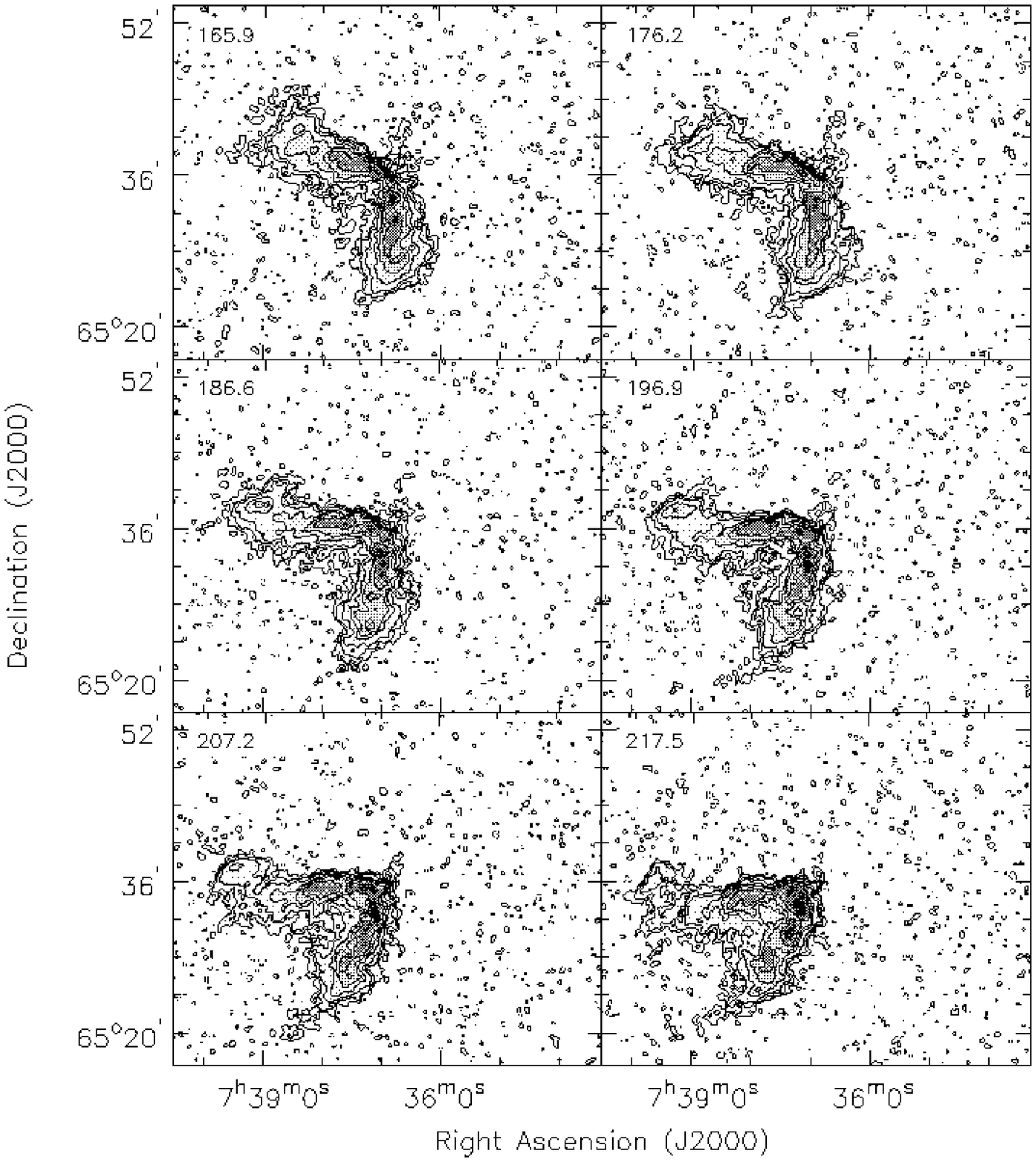}
Figure 2.2 (continue)
\end{figure}

\clearpage
\begin{figure}
\includegraphics[width=\textwidth]{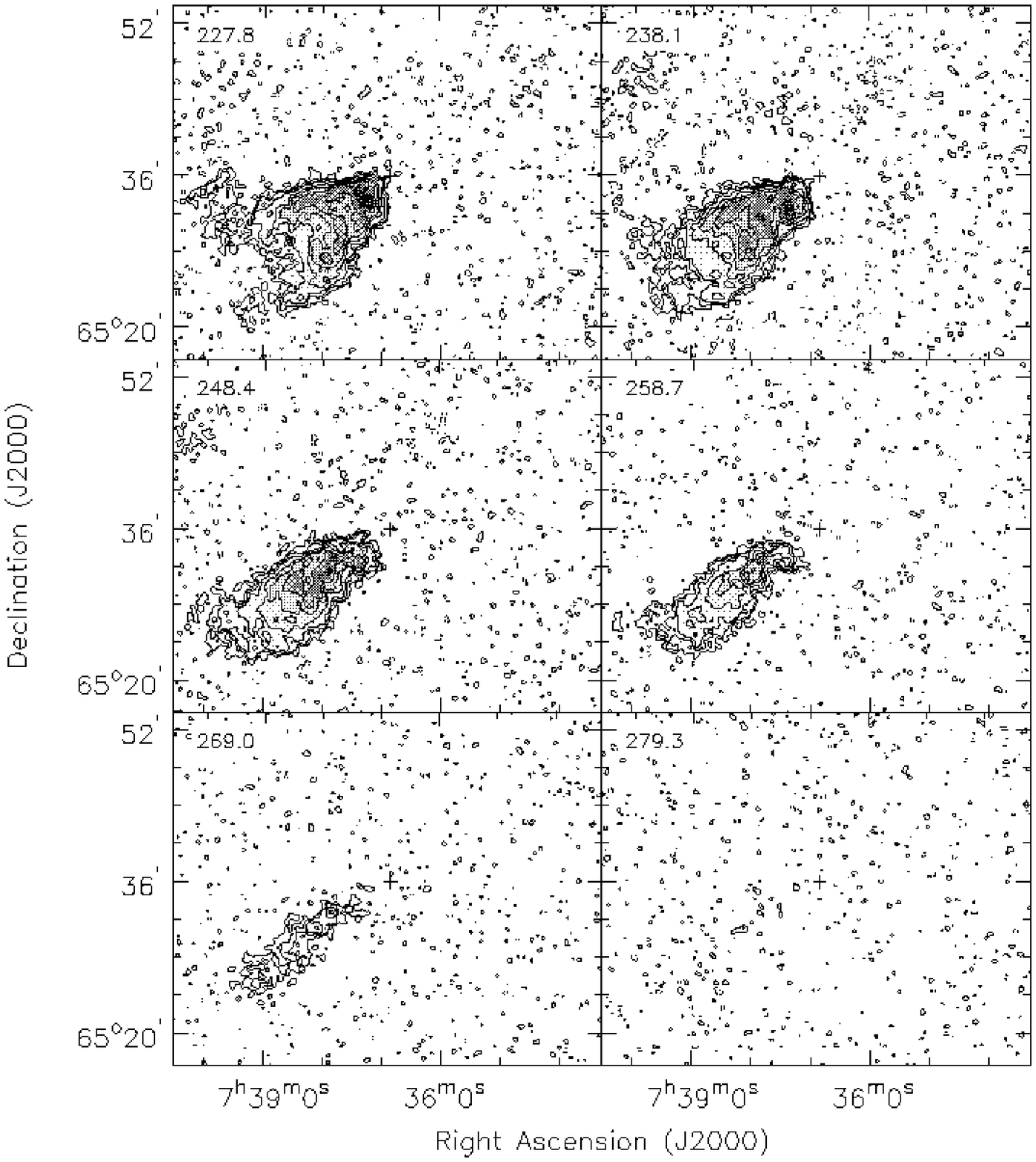}
Figure 2.2 (continue)
\end{figure}

\begin{figure}
\includegraphics[width=120mm]{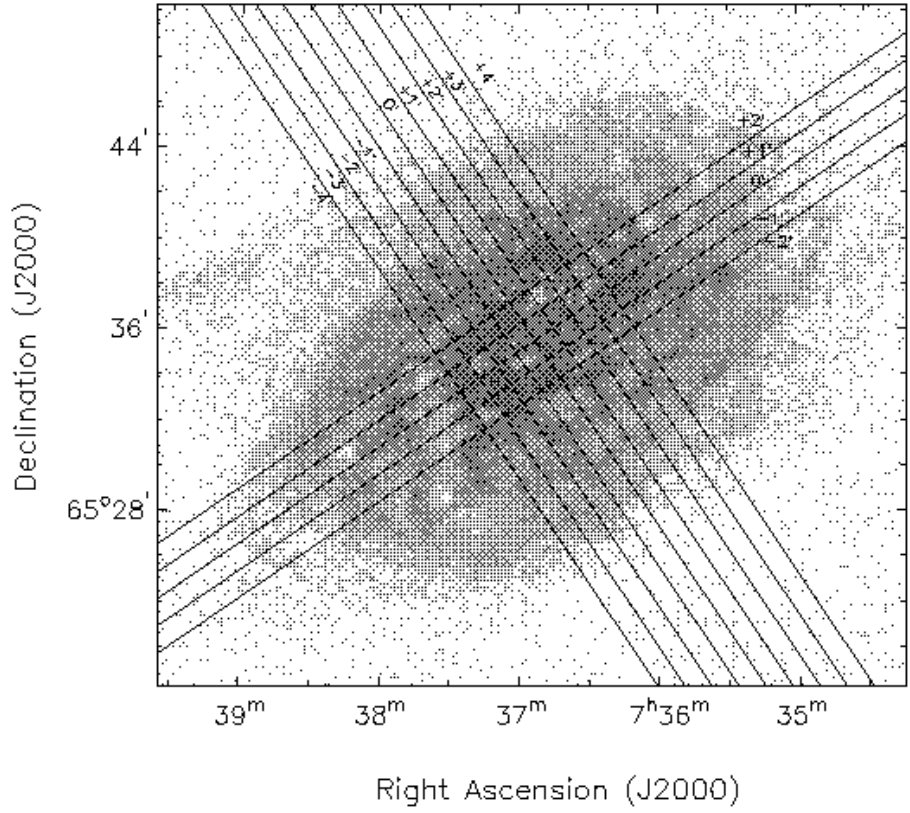}
\figcaption{Total H~{\small I} map showing the
cuts parallel to the major and the minor axes used for the position-velocity 
 maps displayed in Figures~4 and 5.
\label{fig3}}
\end{figure}

\clearpage
\begin{figure}
\includegraphics[width=130mm]{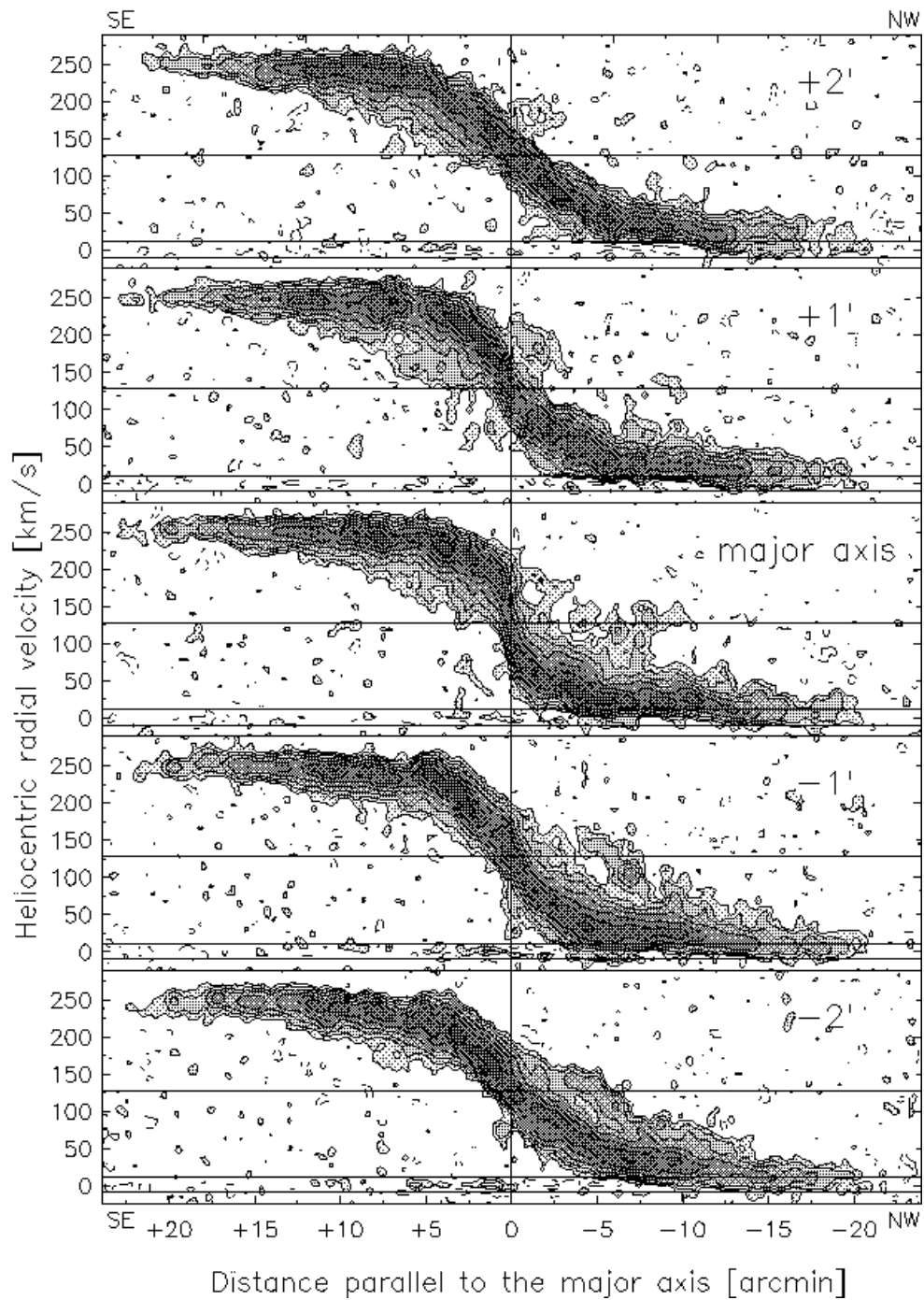}
\figcaption[]{Position-velocity diagrams (30$''$ resolution, Hanning smoothed) 
   parallel to the major axis of
NGC~2403 (p.a.=124.5$^{\circ}$, see Figure~3).
The central panel is the p-v plot along the major axis itself, 
  the +1$'$, +2$'$ are from cuts taken North-East respectively at 1$'$ and 2$'$ from
the centre of the galaxy, the $-$1$'$, $-$2$'$ are located South-West (1$'$
is $\sim$~1 kpc).
The left side of the diagram corresponds to the South-East side of the galaxy.
The central horizontal line shows the systemic velocity, the two lower lines
mark the channel maps contaminated by emission from the Milky Way.
Contours are: $-$0.45, 0.45, 1, 2, 4.5, 10, 20, 45 mJy/beam. 
\label{fig4}}
\end{figure}

\clearpage
\begin{figure}
\includegraphics[width=130mm]{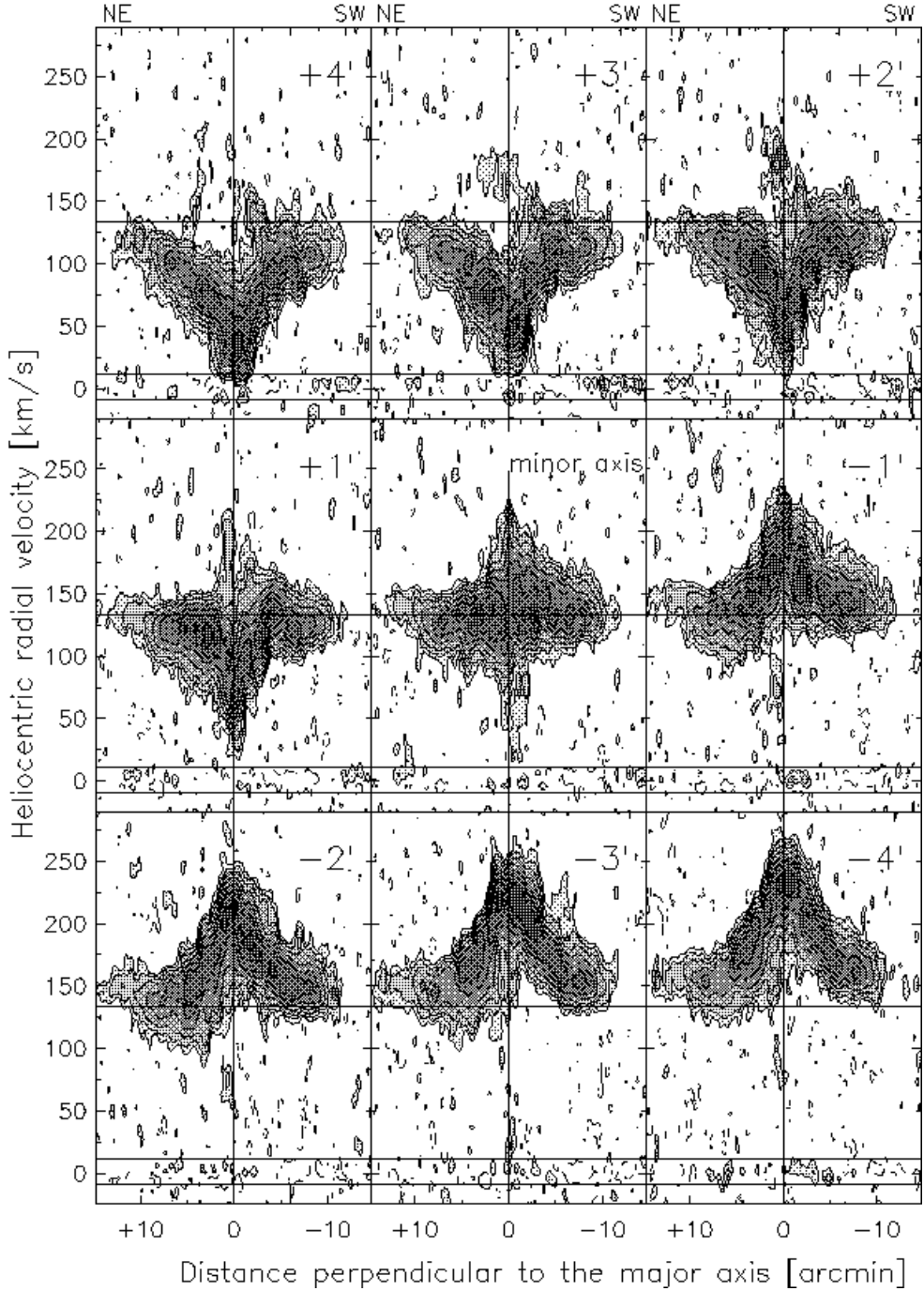}
\figcaption[]{Position-velocity diagrams (30$''$ resolution, Hanning smoothed) 
  parallel to the minor axis of NGC~2403 (p.a.=34.5$^{\circ}$, see Figure~3).
The central panel is the p-v plot along the minor axis, the +1$'$,
+2$'$... are from cuts taken North-West respectively at 1$'$ and 2$'$... from the
centre of the galaxy, the $-$1$'$, $-$2$'$... are located South-East.
The left side of the diagram corresponds to the North-East side of the galaxy.
The central horizontal line shows the systemic velocity, the other two
horizontal lines mark the channels contaminated by H~{\small I} emission from
the Milky Way.
Contours are: $-$0.45, 0.45, 1, 2, 4.5, 10, 20, 45 mJy/beam. 
\label{fig5}}
\end{figure}

\clearpage
\begin{figure}
\plotone{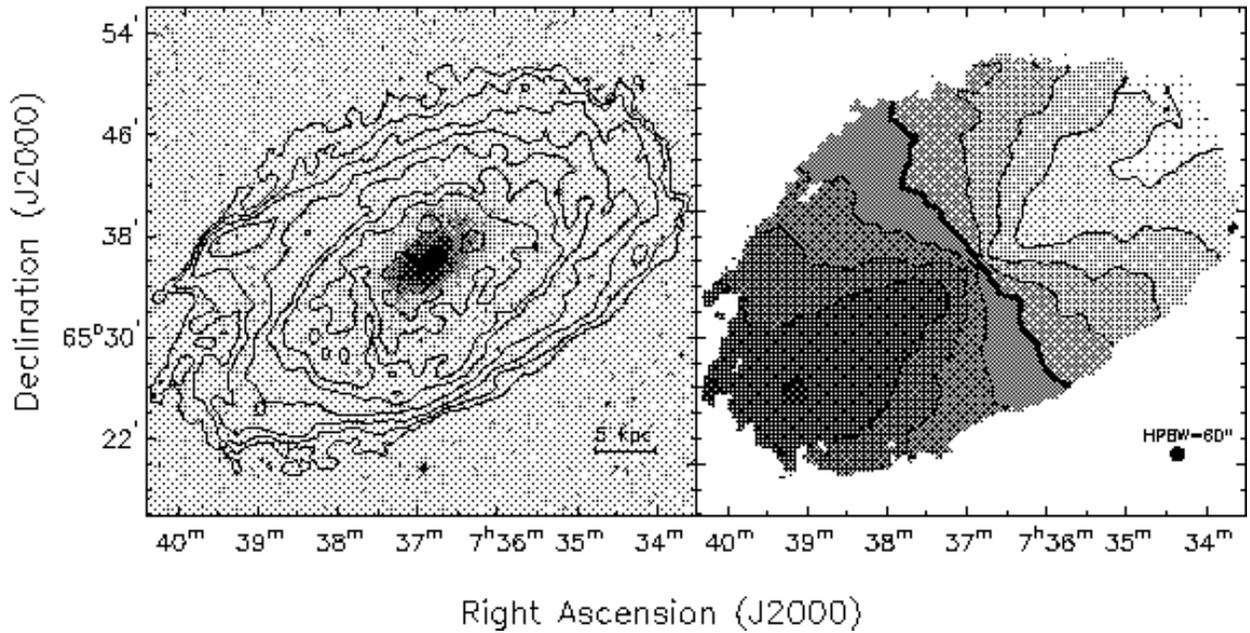}
\figcaption{Total H {\small I} overlaid on the DSS optical image and velocity
field for NGC~2403 at low (60$''$) angular resolution.
The presence of the outer warp is visible on the N-E side of the disk.
The contours are 2.4, 6.4, 13, 26, 52, 78, 126, 207 $\times$10$^{19}$ cm$^{-2}$
(the lower level is $\approx$~2 in terms of r.m.s.\ noise).
In the velocity field the contours are separated by
30 km~s$^{-1}$ and the thicker one marks the systemic velocity (133
km~s$^{-1}$).
The receding side is darker.
\label{fig6}}
\end{figure}

\clearpage
\begin{figure}
\includegraphics[width=130mm]{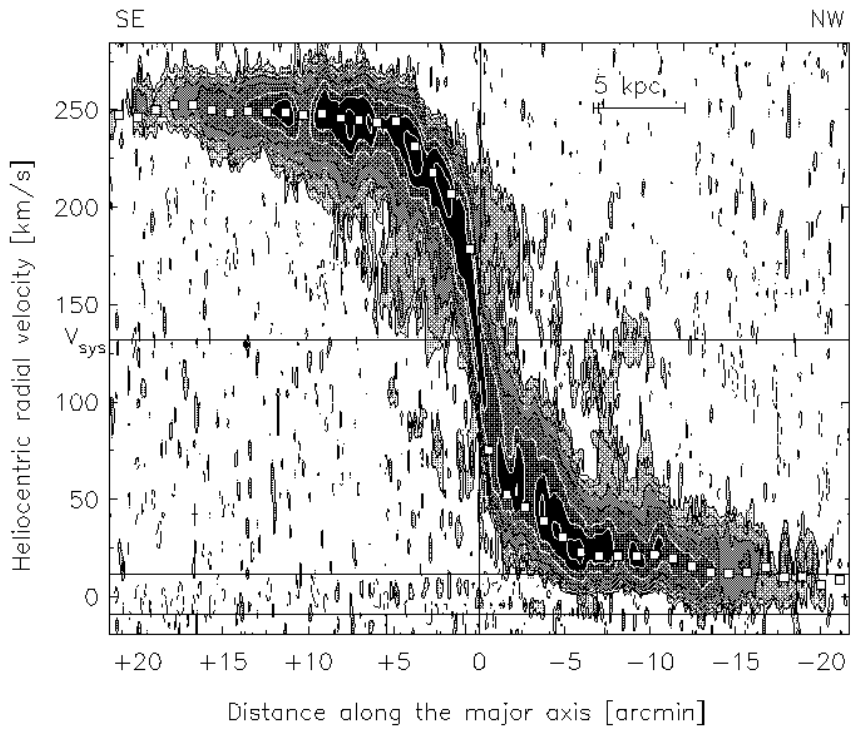}
\figcaption[]{Position velocity diagram (slice $\sim$~1$'$ wide) along
the major axis (p.a.=124.5$^{\circ}$) of NGC~2403.
The spatial resolution is 15$''$, the velocity resolution 10.3 km~s$^{-1}$.
The central horizontal line shows the systemic velocity, the other
two horizontal lines mark the channels contaminated by 
  H~{\small I} emission from the Milky Way.
Contours are: $-$0.26, 0.26, 0.5, 1, 2, 5, 10, 20 mJy/beam.
The r.m.s.\ noise is 0.17 mJy/beam.
White squares mark the rotation curves for the two sides of the galaxy.
\label{fig7}}
\end{figure}

\clearpage
\begin{figure}
\plotone{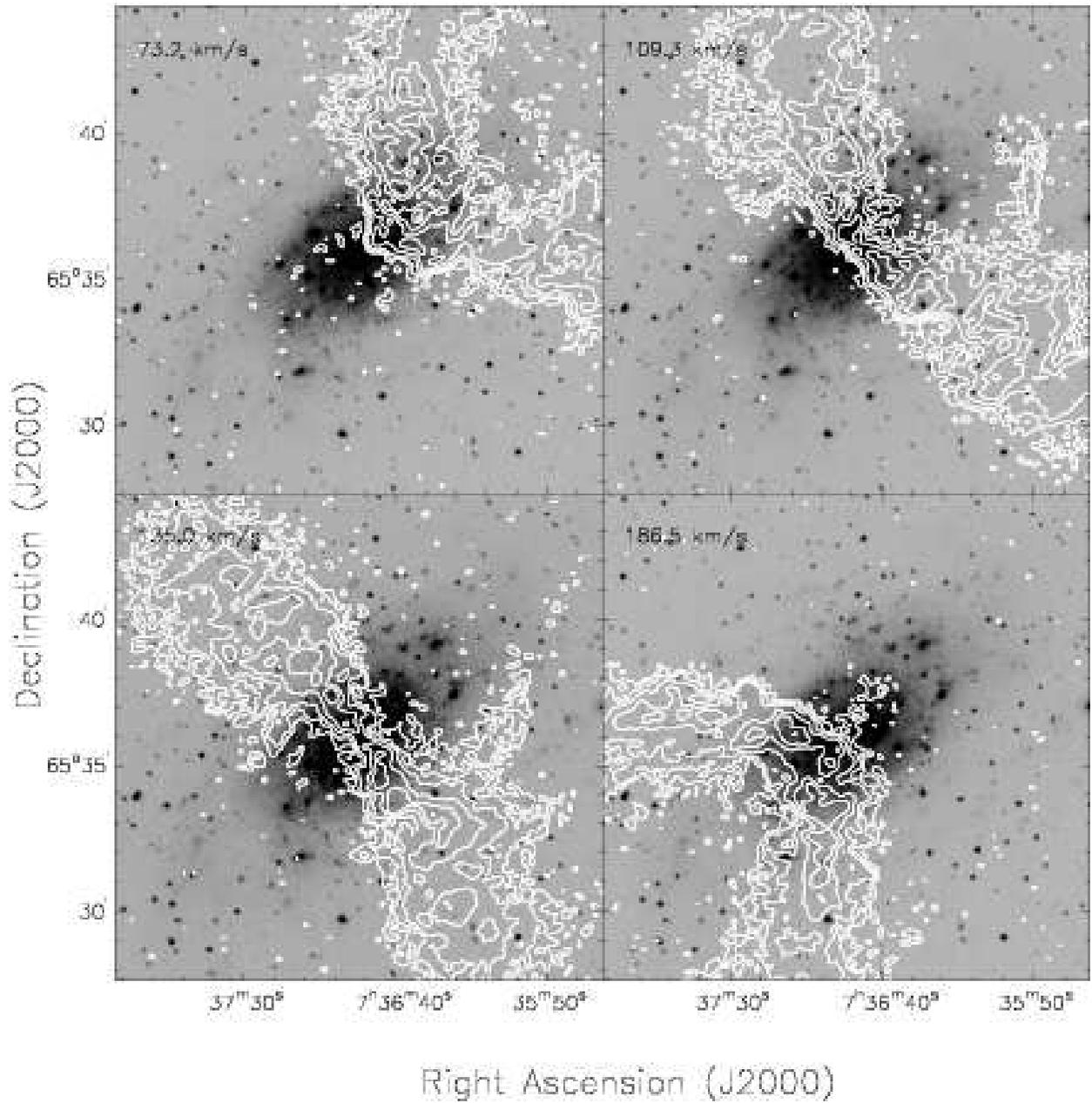}
\figcaption[]{Four representative channel maps at 15$''$ resolution
overlapping the optical image of NGC~2403.
Contour levels are: $-$0.5, 0.5, 1, 2, 5, 10, 20
mJy/beam. The r.m.s.\ noise is 0.17 mJy/beam.
The lower right panel shows the forbidden gas in the North-West side of
the galaxy while the upper left panel shows the faint emission of
forbidden gas in the South-East side.
The other panels show the bright 8 kpc filament in the West side of the H~{\small I} disk.
Note its location outside the bright part of the optical disk.
\label{fig8}}
\end{figure}

\clearpage
\begin{figure}
\plotone{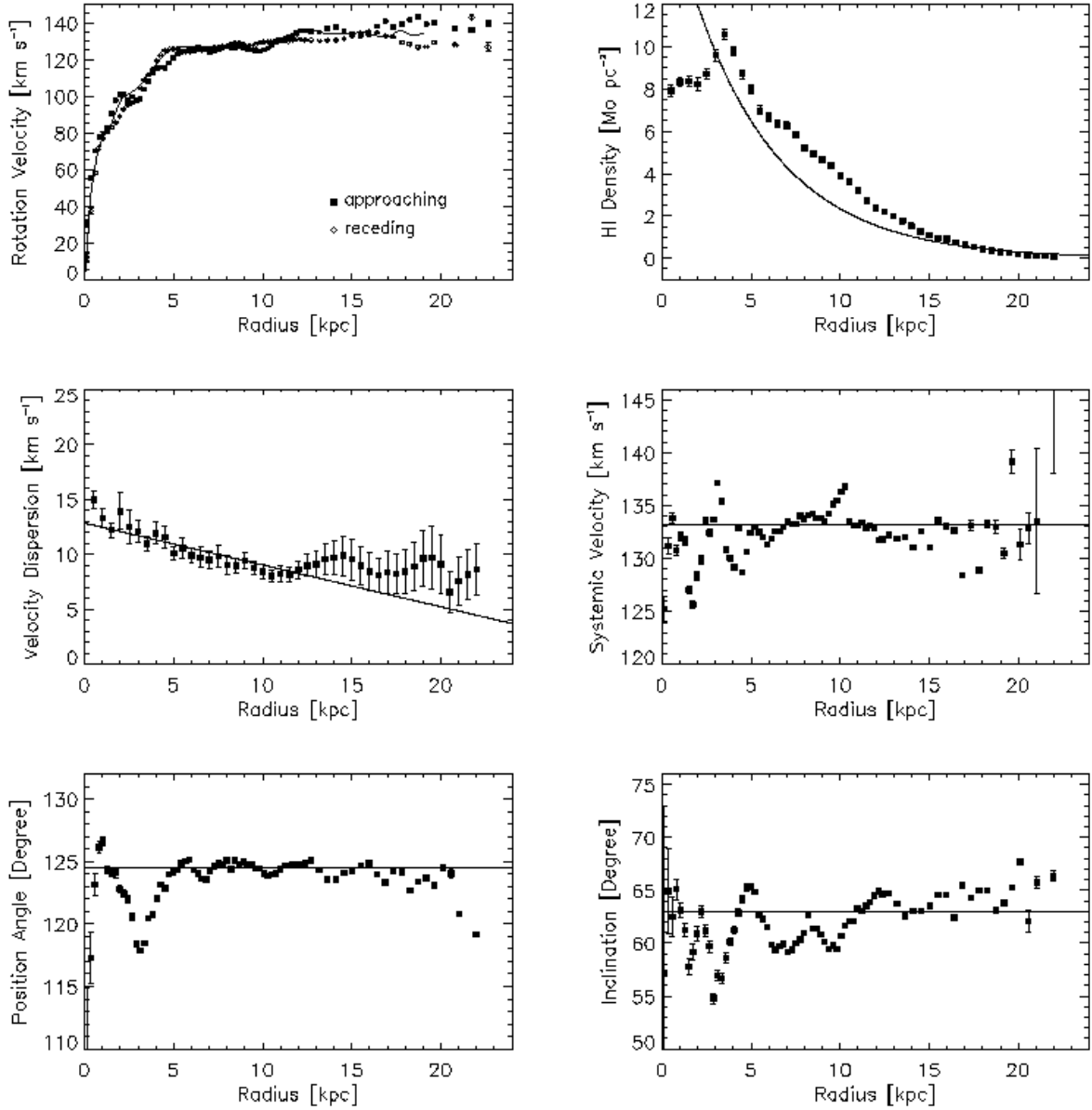}
\figcaption[]{Radial profiles of parameters of the tilted ring model for
NGC~2403.
In the top left panel the rotation curves obtained for the two sides of
the galaxy are shown.
The line shows the rotation curve obtained by Begeman (1987).
\label{fig9}}
\end{figure}

\clearpage
\begin{figure}
\plotone{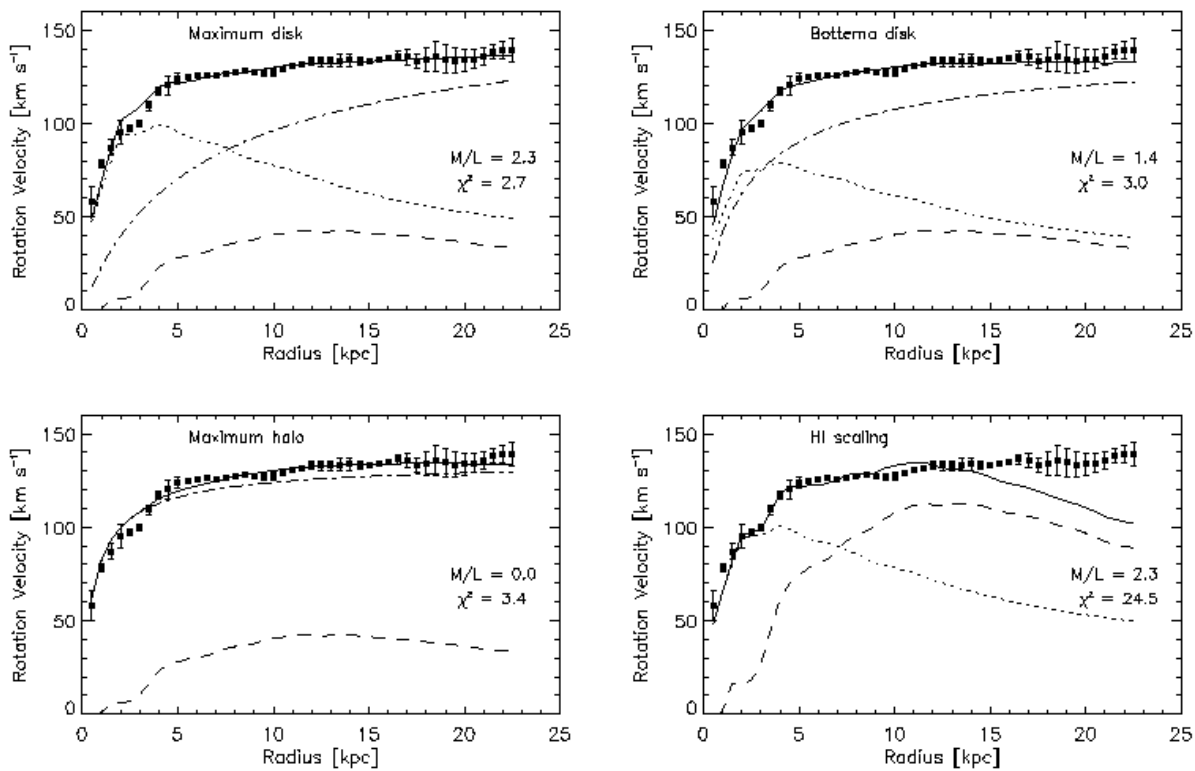}
\figcaption[]{Four different mass models for NGC~2403.
The filled squares show the observed rotation curve,
the contributions by gas (dashed), stars (short dashed), and DM
halo (long-short dashed) and the total rotation velocity (solid).
Values of $\chi^2$ and scaling factor (M/L) of the stellar profile are
given as well.
\label{fig10}}
\end{figure}

\clearpage
\begin{figure}
\plotone{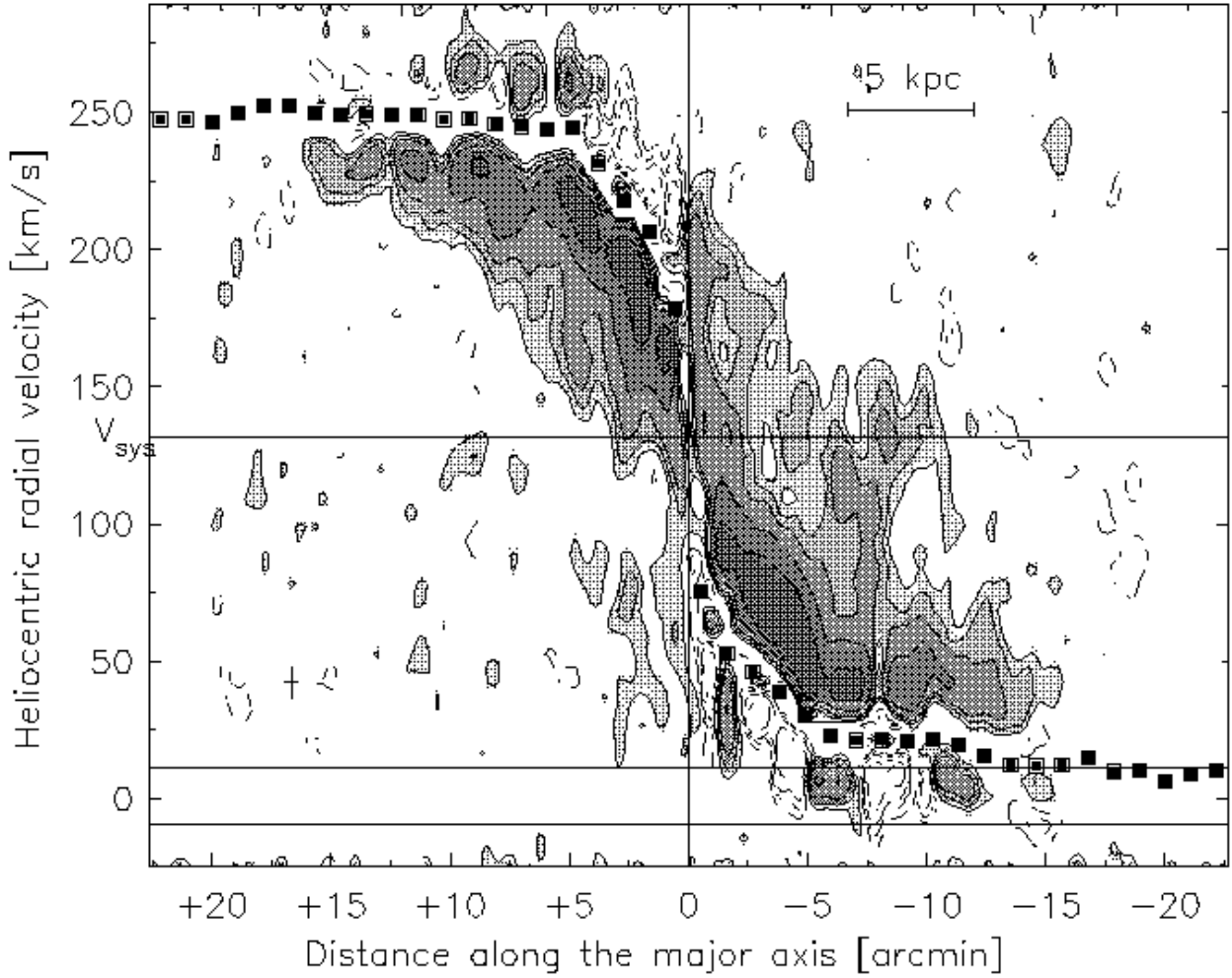}
\figcaption[]{Position-velocity diagram for the anomalous gas along the
major axis of NGC~2403. Contours are 2, 4, 8, 20, 40 in units of r.m.s.\
noise.
The filled squares mark the rotation curve of the cold disk.
\label{fig11}}
\end{figure}

\clearpage
\begin{figure}
\plotone{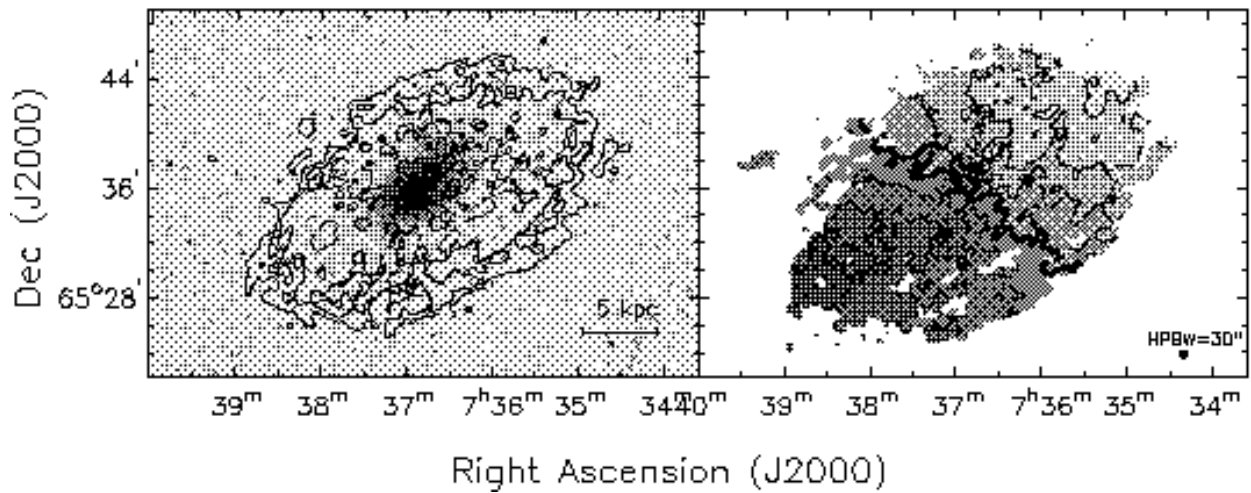}
\figcaption[]{Total H~{\small I} map and weighted mean velocity field of the
anomalous gas in NGC~2403. 
In the left panel the density distribution of the anomalous gas is overlaid on
the optical DSS image of the galaxy.
The contours are 2, 6, 15 in units of r.m.s.\ noise.
In the velocity field (right) the contours are separated by 
30 km~s$^{-1}$ and
the thick one shows the systemic velocity (kinematical minor axis).
Note that this latter is rotated counter-clockwise (by about 20 degrees) with
respect to the minor axis of the velocity field of the cold disk shown in
Figure 1. 
\label{fig12}}
\end{figure}

\clearpage
\begin{figure}
\plotone{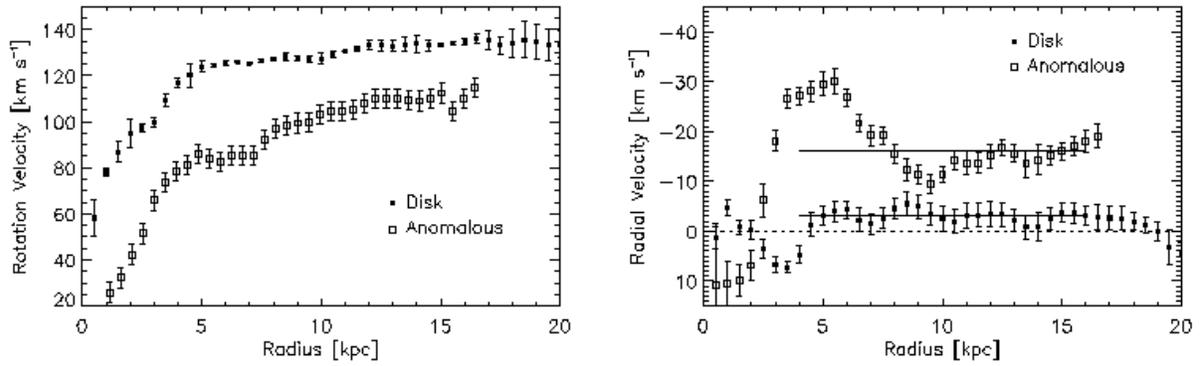}
\figcaption[]{Rotation curves (left) and radial curves (right) for the 
the cold disk (filled squares) and for the  anomalous gas 
(open squares).
The mean values of the radial inflow velocity for the anomalous gas and 
for the disk (lines) are respectively 16.2 km~s$^{-1}$ 
and 3.1 km~s$^{-1}$.
\label{fig13}}
\end{figure}

\clearpage

\begin{deluxetable}{lc}
\tabletypesize{\scriptsize}
\tablecaption{Observational parameters for NGC~2403. \label{tab1}}
\tablewidth{0pt}
\tablehead{
\colhead{} &
\colhead{NGC~2403}
 }
\startdata
Observation dates               &  Jan~1,~28,~31; Feb~1,~1999 \\
Length of observation (hours)   &  48\\
Time on source (hours)          &  40\\
Configuration                   &  $\mathrm CS$\\
Number of antennas              &  27\\
Pointing R.A. (J2000)                   &  07 36 54.5\\
Pointing Dec  (J2000)                   &  +65 35 20.0\\
Central velocity (km~s$^{-1}$) &  130\\
Central frequency (MHz)         &  1419.747\\
Total bandwidth (MHz)           &  3.125\\
Total bandwidth (km~s$^{-1}$)  &  660\\
Number of channels              &  127\\
Channel separation (kHz)        &  24.4\\
Channel separation (km~s$^{-1}$)& 5.1\\
\enddata
\end{deluxetable}

\clearpage

\begin{deluxetable}{lcc}
\tabletypesize{\scriptsize}
\tablecaption{Parameters of the data cubes. \label{tab2}}
\tablewidth{0pt}
\tablehead{
\colhead{} & \colhead{Full resolution} & \colhead{Smoothed}
}

\startdata
Used value of ROBUST    & 0.2                           & 0\\
Used taper ($''$)       & no taper                      & 27\\
HPBW  ($''$)            & 15.2 $\times$ 13.3            & 29.7 $\times$ 29.3\\
P.A. of synthesized beam ($^{\circ}$)& 91.7             & 98.7\\
Beam size (pc)          & 234 $\times$ 205              & 458 $\times$ 452\\
Number of channels      & 62                            & 62\\
R.m.s.\ noise per channel:&                              & \\
~~~~~No Hanning smoothing (mJy/beam)  & 0.28            & 0.35\\
~~~~~Hanning smoothing (mJy/beam)     & 0.17            & 0.22\\
Minimum detectable column density: &                    &\\
~~~~~(cm$^{-2}$) & 5.0 $\times$ 10$^{19}$ & 2.0 $\times$ 10$^{19}$\\
~~~~~($M_{\odot}$/pc$^2$)    &   0.38                   & 0.15\\
Minimum detectable mass ($M_{\odot}$/beam)&2.1 $\times$ 10$^4$&2.6 $\times$10$^4$\\
Conversion factor mJy/K & 3.0                           & 0.7\\
\enddata
\end{deluxetable}

\clearpage

\begin{deluxetable}{lcc}
\tabletypesize{\scriptsize}
\tablecaption{Optical and radio parameters for NGC~2403. \label{tab3}}
\tablewidth{0pt}
\tablehead{
\colhead{} &
\colhead{NGC~2403} &
\colhead{ref}
}
\startdata
Morphological type      & Sc(s)~III                      & 1 \\
Optical centre ($\alpha$, $\delta$ J2000)  & 7~36~51.92 ~~65~36~0.6         & 2\\
Kinematical centre ($\alpha$, $\delta$ J2000)& 7~36~50.66 $\pm$ 1.48 ~~65~36~2.2 $\pm$ 5.5 & 3\\
Distance (Mpc)          & 3.18                           & 4 \\
L$_B$   (${L_{\odot}}_B$)       & 7.9 $\pm$ 0.7 $\times$ 10$^9$ & 5\\
Disk scale length (kpc) & 2.0                            & 6\\
R$_{25}$ (kpc)          & 8.2                            & 5\\
Holmberg radius (kpc)   & 13.4                           & 7\\
H~I scale length (kpc)   & 5.7 $\pm$ 0.1                  & 3\\
H~I disk radius  (kpc)   & 22.5                             & 3\\
Systemic velocity (km~s$^{-1}$)& 133.2 $\pm$ 2.2         & 3\\
H~I total mass ($M_{\odot}$)  & 3.24 $\pm$ 0.05 $\times$ 10$^9$ & 3\\
Mean H~I inclination ($^{\circ}$)        & 62.9 $\pm$ 2.1      & 3\\
Mean P.A. (no warp) ($^{\circ}$)        & 124.5 $\pm$ 0.6     & 3\\
Total mass ($M_{\odot}$)& 9.5 $\pm$ 0.7 $\times$ 10$^{10}$    & 3\\
\enddata

\tablerefs{
(1) \citet{san81}
(2) \citet{gal73};
(3) this work;
(4) \citet{mad91};
(5) \citet{dev76};
(6) \citet{wev86};
(7) \citet{hol58}.
}
\end{deluxetable}

\clearpage

\begin{deluxetable}{lcc}
\tabletypesize{\scriptsize}
\tablecaption{Comparison between disk and anomalous gas. \label{tab4}}
\tablewidth{0pt}
\tablehead{
\colhead{} & \colhead{Cold disk} & \colhead{Anomalous gas}
}
\startdata
Mass ($M_{\odot}$)       & 3 $\times$ 10$^9$       & 3 $\times$ 10$^8$ \\
Diameter (kpc)  & 46                    & 32\\
Thickness (kpc)      & 0.4           & $<$3\\
Maximun rotation velocity (km~s$^{-1}$)& 130            & 110\\
Mean radial inflow velocity (km~s$^{-1}$)& 0$-$3           & 10$-$20\\
Mean velocity dispersion (km~s$^{-1}$) & 8-12           & 20$-$50\\
\enddata
\end{deluxetable}

\end{document}